%% file: main.tex
\documentclass{misc/IEEEcsmag}
\usepackage{subcaption}
\usepackage{color}
\usepackage{xcolor}
\usepackage{tikz}
\usepackage[hidelinks]{hyperref}
\usepackage{orcidlink}
\usepackage{todonotes}
\let\labelindent\relax
\usepackage{enumitem}

\definecolor{reda}{rgb}{1.0, 0.44, 0.37}

\DeclareRobustCommand{\blackcircled}[1]{%
  \tikz[baseline=(char.base)]{
    \node[shape=circle, fill=black, inner sep=1pt] (char)
      {\textcolor{white}{\small \textbf{#1}}};
  }%
}

\jvol{XX}
\jnum{XX}
\paper{8}
\jmonth{February}
\jname{IEEE Internet Computing}
\pubyear{2026}

\begin{document}

\sptitle{DEPARTMENT: INTERNET OF THINGS, PEOPLE, AND PROCESSES}

\title{Service Orchestration in the Computing Continuum: Structural Challenges and Vision}

\author{Boris Sedlak~\orcidlink{0009-0001-2365-8265}, Víctor Casamayor Pujol~\orcidlink{0000-0003-2830-8368}, Ildefons Magrans de Abril~\orcidlink{0000-0002-1815-5372}}
\affil{Universitat Pompeu Fabra, Barcelona}

\author{Praveen Kumar Donta~\orcidlink{0000-0002-8233-6071}}
\affil{Stockholm University, Stockholm, Sweden}

\author{Adel N. Toosi~\orcidlink{0000-0001-5655-5337}}
\affil{University of Melbourne, Melbourne, Australia}

\author{Schahram Dustdar~\orcidlink{0000-0001-6872-8821}}
\affil{ICREA, Barcelona, Spain; and TU Wien, Vienna, Austria}

% \author{Might ask others}
% \affil{...}
% Like Adel N. Toosi, Anastasios Zafeiropoulos, Juanma or someone from his lab (e.g., Sergio)

\markboth{INTERNET OF THINGS, PEOPLE AND PROCESSES}{INTERNET OF THINGS, PEOPLE AND PROCESSES}

\begin{abstract}
% \looseness-1 
    % To minimize latency in Internet of Things (IoT) applications, computing has shifted from the Cloud to resource-limited Edge devices, leading to the Computing Continuum (CC). While the CC enhances efficiency, it complicates service orchestration due to heterogeneous device capabilities. To simplify the orchestration of CC systems, we (1) improve service reactivity through flexible behavioral models, which we (2) maintain accurate through continuous interaction with the environment. Finally, we (3) optimize global performance in the CC by analyzing the dependencies between services and hosting devices.
    The Computing Continuum (CC) integrates different layers of processing infrastructure---from Edge to Cloud---to optimize service quality through ubiquitous and reliable computation. Compared to central architectures, however, heterogeneous and dynamic infrastructure increases the complexity for service orchestration.   %
    To guide research, this article first summarizes structural problems of the CC, and then, envisions an ideal solution for autonomous service orchestration across the CC.
    As one instantiation, we show how Active Inference---a concept from neuroscience---can support self-organizing services in continuously interpreting their environment to optimize service quality.
    Still, we conclude that no existing solution achieves our vision, but that research on service orchestration faces several structural challenges.
    % Still, neither this instantiation nor existing solutions fully address the structural challenges of CC service orchestration.
    Most notably: provide standardized simulation and evaluation environments for comparing the performance of orchestration mechanisms.
    % (1) simulation and benchmarking frameworks for reproducible training and evaluation, (2) causal and explainable reasoning to support trustworthy orchestration, (3) cross-vendor and personal infrastructure integration to preserve data sovereignty, and (4) automation and antifragility to ensure resilient, self-improving operation.
    Together, the challenges outline a research roadmap toward resilient and scalable service orchestration in the CC.
\end{abstract}
\maketitle

% \TODO{@Contributors: Please check (1) if the challenges make sense from a logical point of view, (2) if yes, how the presentation could be improved (e.g., figures or formulas), (3) how/where to link the challenges better to the initial approach, and (4) if you'd like to extend the challenges.}

% 6,000 words including all text, the abstract, keywords, bibliography, biographies, and 250 words for each figure and table.

\section{INTRODUCTION}

Sensory data from Internet of Things (IoT) devices form the backbone of pervasive applications. As running example, consider a digital twin of an entirely smart city that allows clients to interact with content through Augmented Reality (AR) across surfaces, like handhelds or windows. To enter and navigate such environments in real-time, computation is shifted from Cloud centers towards Edge devices; thus, it is possible to process and render content on nearby devices. While this reduces latency, Edge devices offer limited and less predictable resources.
During environmental changes, e.g., when facing higher load at rush hours, Edge devices require fallback mechanisms that ensure service quality.
To combine the strengths of both worlds, Cloud and Edge layers are integrated in one composite architecture---the Computing Continuum (CC). Thus, applications can place latency-aware services on the Edge, and use the abundance of Cloud resources for hosting the remaining services.

% While the CC's impact on service-oriented computing is compelling---mostly due to improved Quality of Experience (QoE)---it introduces numerous research challenges
While the CC aims to improve Quality of Experience (QoE), it introduces numerous orchestration challenges~\cite{plebani_service-oriented_2024}.
Most notably, by distributing services from one application across different physical devices with heterogeneous characteristics, it gets complex to predict application behavior (e.g., after scaling up one service).
%
% This is further complicated because CC applications are likely distributed over devices from different vendors, or across jurisdictions. While the goal is optimizing the service quality across such borders, vendors might not be willing to share their device and service states, or only partially.
This is further complicated by deploying applications across multiple vendors and jurisdictions, as providers may be unwilling to share their full system state.
This demands solutions that allow parties with partial observability to collaborate towards common optima, while accounting for their individual behavior. 

Driven by the numbers of IoT devices and the need to process data nearby, the last decade has produced a large amount of research to cope with the heterogeneity and distribution of processing systems. Common topics revolve around latency-aware service placement~\cite{lan_sla-orecs_2024}, or forming agreements between service and device providers~\cite{kochovski_smart_2020}. These mechanisms are often based on static architectures and require a priori understanding of services and devices, e.g., benchmarking the service quality across device types. However, individual devices in the CC will (dis)appear dynamically, which can include new device types that are not benchmarked yet. Also, clients can always redefine the desired service operation---usually specified through \textit{Service Level Objectives} (SLOs)---Thus putting the service in an unknown context.
To summarize these problems, this article provides a structured overview of hypotheses that describe the CC, its applications, and challenges for service orchestration.

To guide research on service orchestration---also inspired by the well-known \textit{vision of autonomic computing}~\cite{kephart_vision_2003}---this article outlines how continuous service interpretation and adaptation can help optimize SLO fulfillment.
To instantiate this design, we present a preliminary implementation using Active Inference (AIF)~\cite{parr_active_2022}---a concept from neuroscience that aims to create self-organizing components that maintain internal requirements fulfilled: 
First, we model the interactions between a service and its environment through a behavioral Markov blanket (MB)---a description of \textit{how} a service interprets its current state and \textit{which} corrective action to take~\cite{sedlak_designing_2023_short}. Thus, services can decide how to react depending on the context, e.g., by shifting computation or services accordingly. Second, whenever the context changes, e.g., after dynamically reconfiguring SLOs, these behavioral models must be updated. Therefore, we wrap each component in a continuous action-perception cycle and adjust its MBs according to environmental feedback~\cite{sedlak_equilibrium_2024}. Third, we analyze interactions between services, and additionally, their hosting devices, by composing their MBs~\cite{sedlak_markov_2024_short}. This enhances collaboration within the CC because services can estimate how local actions impact dependent services and the corresponding resource demand. 

This implementation, using AIF, allows collaborative agents to continuously model and understand the environment---a first step to address fundamental problems of dynamic CC systems. However, when mapping the implementation to the structural hypotheses and problems, we still identify multiple challenges that impede research on service orchestration. Most notably, we see a clear gap for: (1) large-scale, standardized simulation environments that simplify testing hypotheses and comparing solutions, (2) continuous and context-aware mechanisms for accurate ML inference, and (3) seamless infrastructure composition that allows clients to use infrastructure from multiple vendors or individuals. To achieve structural improvement for service computing, we summarize these challenges in more detail under three key challenge areas. 

%
% To achieve structural improvement for service computing in general, this paper creates 
% We wrap these challengs in three groups  that promise structural improvement---not only for our solution---but service computing in general.

\section{STRUCTURAL DEFINITION}

Research on CC systems is in an early stage, resulting in numerous beliefs and views about their inherent challenges, like in~\cite{kochovski_smart_2020,sedlak_equilibrium_2024,akbari_icontinuum_2024}. In this section, we provide a structural perspective to orchestration problems in CC systems and design a general guideline for solutions to them. 
Our core objective: create a holistic solution that optimizes processing requirements throughout CC applications; in practice, this means capturing the desired state through SLOs (e.g., expected latency or quality) and making them first-class citizens during the entire system operation.
This abstract solution invites and permits different instantiations---the one provided by us in the next section, being one of them.

% \rw{Leveraging the proposed structure aims at enabling the following objectives: 
% \textbf{Generality}. Support different types of distributed applications; \textbf{Composability}. Allow the combination of multiple decision-makers; \textbf{Adaptivity}. Enable continuous adaptation to workload and resource changes; 
% \textbf{SLO-awareness}. Make SLO first-class citizens in the decision process; \textbf{CC-awareness}. Exploit the presence of multiple computing layers; \textbf{Stability}. Ensure stability of the closed-loop behavior under variable workloads and uncertainties.;
% \textbf{Extensibility}. Allow plugging new decisions or architectures without changing the rest of the structure.}

\subsection{Problem Hypotheses}

In the following, we provide three sets of hypotheses that define the problem domain, its management, and ways to model this.
First, the \textbf{domain hypotheses}, are assumptions on inherent problems of the CC.

\begin{enumerate}[label=D.\arabic*]
    \item \textbf{Heterogeneous continuum}. The computing continuum spans all computational tiers, implying a high level of heterogeneity in hardware or network capacity, and energy consumption.
    
    \item \textbf{Dynamic conditions}. Resource availability, workload intensity, network conditions, and SLOs from clients will vary over time. This can change at multiple time scales, including fast-paced periodic fluctuations, or slow-paced behavioral drifts.
    
    \item \textbf{Partial and decentralized observability}. Due to distribution, communication delays, and administrative boundaries, no management component can observe the full system state instantaneously.
    
    \item \textbf{Large-scale}. The CC infrastructure creates a highly-distributed, dense network of computing nodes that spans vast geographical areas.
\end{enumerate}

 \vspace{6pt}

Second, we define \textbf{application hypotheses} that define CC services and their management scope.

\begin{enumerate}[label=A.\arabic*]
    
    \item \textbf{Interdependent services}. Applications in the CC are composed of multiple interacting services or microservices, making their performance dependent on upstream services and their resources.
    
    \item \textbf{Multi-tenant and multi-vendor applications}. Applications are composed of services and computing nodes that belong to different parties, so that no entity has control over the entire system.
    
    \item \textbf{SLO-based objectives}. Each application (or service) is associated with Service Level Objectives (SLOs), expressed in relation to measurable metrics, that define the desired application state.
    
    \item \textbf{Service instrumentation}. Each service exposes monitoring data (e.g., metrics, logs, traces) and control interfaces (e.g., scale, migrate, change configuration) to observe and act on services.
\end{enumerate}

\vspace{6pt}

Third, we describe the \textbf{modeling hypotheses}, so which abstraction are valid to describe the problem.

\begin{enumerate}[label=M.\arabic*]
    \item \textbf{State-space representation}. The CC can be described by high-dimension state vectors---$\mathcal{S}$---that comprise resource, service, and SLO states. Still, agents can observe only a subset of the true state.

    \item \textbf{Action-space representation}. The control interfaces exposed by services and devices mark the boundaries of the CC's action space---$\mathcal{A}$. Technically, agents could invoke all types of action remotely; however, we limit their scope and permissions to a hierarchical subset of actions.
    
    \item \textbf{Stochastic dynamics}: State transitions are stochastic---identical actions can lead to different resulting states due to environmental variability. 
    
    % \item \textbf{Multi-scale control}. Considering that dynamic processes operate at different time scales, solutions to this 
\end{enumerate}

% \subsection{Problem boundaries}
% \tmi{Formally, the structural definition concerns the runtime management of applications defined over a state space $\mathcal{S}$, with a set of admissible actions $\mathcal{A}$, subject to infrastructure and policy constraints $\mathcal{C}$.
% %
% The runtime management of distributed applications over the CC, includes: initial placement and migration of services, scaling decisions, configuration adjustment, and routing, while maintaining all application SLOs fulfilled and the resources and policies within the specified constraints.
% %
% The structural solution shall support decision making at multiple layers of the CC.
% %
% The logic of the management actions operates in recurrent control loops, this might be periodic, event-driven or a hybrid of both.
% %
% The structural solution shall allow centralized, hierarchical, and fully distributed controllers.}

% \subsection{Problem Boundaries}

% Under these hypothesis, we can design solutions that dynamically orchestrate applications and services at runtime, e.g., by scaling or reconfiguring services.
% %
% Formally, this state space $\mathcal{S}$, with a set of admissible actions $\mathcal{A}$, subject to infrastructure and policy constraints $\mathcal{C}$.

Under these hypotheses, we can design solutions that autonomously manage applications within the CC, e.g., by scaling or reconfiguring services to fulfill SLOs.
% In the following, we outline one possible way to address these problems across the CC through an omnipresent adaptation cycle.
In the following, we envision an abstract solution design for service orchestration in the CC that builds upon the well-known \textit{vision of autonomic computing}~\cite{kephart_vision_2003}.

\subsection{Abstract Solution Design}

To optimize service orchestration in the CC, we provide an abstract design guideline split into two parts: a high-level agentic lifecycle for monitoring and optimizing processes, and an instantiation of this cycle using custom tools and mechanisms. By following this agentic cycle, intelligence can be created at a desired scale, thus interlacing the CC with self-adaptive components. First, regardless of the infrastructure or application, smart components must implement a closed-loop architecture~\cite{kephart_vision_2003}; this forms the \textbf{invariant} solution part:

% \begin{enumerate}[label=I.\arabic*)]
%     \item \textbf{Observation} component. Collects and preprocesses monitoring data to infer parts of $\mathcal{S}$.
    
%     \item \textbf{Knowledge} component. Maintains system states $\mathcal{S}$, their dynamics and relations to SLOs.
    
%     \item \textbf{Decision-making} component. Infers actions $a \in \mathcal{A}$ based on local knowledge and SLOs.
    
%     % \item \textbf{Actuation} component. Enacts decisions on the underlying infrastructure and applications.
    
%     \item \textbf{Coordination} component. Manages interactions between multiple decision makers (e.g., negotiation, conflict resolution, hierarchy).
% \end{enumerate}

% Based on the established components, the system operates through a closed-loop architecture. 
% The following information flows are invariant:

\begin{itemize}
    \item Observation $\rightarrow$ Knowledge: Monitoring data is collected and interpreted to transform observations $o_t$ to internal state representations $\hat{s}_t$; for example, monitor CPU load to be aware of device utilization.
    
    \item Knowledge $\rightarrow$ Decision-making: an action $a_t \in \mathcal{A}$ is inferred based on the state $\hat{s}_t$, an internal knowledge model $\mathcal{M}$, and SLO preferences $\mathcal{O}$; for example, note excessive device utilization, and infer that offloading computation will improve latency.
    
    \item Decision-making $\rightarrow$ Actuation: the action $a_t$ is executed in the CC's infrastructure and applications. For example, scaling an application horizontally.
    
    \item Actuation $\rightarrow$ Observation: actions modify the system state, generating new observations $o_{t+1}$.
    
    \item Coordination $\leftrightarrow$ all components: exchange summaries $\xi_t$ of states, intents, and constraints to improve coordination between decision makers; for example, using knowledge transfer or distillation.
\end{itemize}
Any control mechanism for the CC must implement these steps in a recurrent loop of the form
\[
o_t \rightarrow \hat{s}_t \rightarrow a_t \rightarrow o_{t+1}.
\]

While the structure of autonomous control loops is invariant, the individual steps can be instantiated differently to create customized implementations of agents; this forms the \textbf{variant} part of the solution:

\begin{enumerate}[label=V.\arabic*]
    \item \textbf{Architecture}. Centralized or distributed agents with different hierarchical depth or granularities, e.g., agents per service, application, region etc.
    
    \item \textbf{Configuration}. Which actions and states are available at each control layer, how often are control loops (action-perception) executed, and what operational overhead is tolerable per agent.
    
    \item \textbf{Modeling}. Design the state representation purely metric-based, or graph-based with dependencies. Keep probabilistic beliefs about hidden states or directly observe states. Model state and action spaces as discrete or continuous variables.
    % --> though the question if it's directly observable depends on the domain
    
    \item \textbf{Algorithm}. Decision models can be based on heuristic rules, numerical optimizers, active inference, (multi-agent) RL, or other techniques.
    
    % \item \tmi{\textbf{Constraint handling}. Instantiations may incorporate different constraint representations and solvers (e.g., linear constraints, probabilistic constraints, safety envelopes).}
\end{enumerate}

~

Based on the presented hypotheses and the abstract solution, the next section shows a custom implementation for parts of this abstract solution.

\section{PRELIMINARY IMPLEMENTATION}

% \ins{storyline: we present our framework that addresses the fundamental service-computing-related challenges from the introduction. We highlight the implications of these contributions and pinpoint future points of improvement; this we do in a Discussion/Limitation subsection }

\begin{figure*}
\vspace{-20pt}
    \centering
    \includegraphics[width=0.95\linewidth]{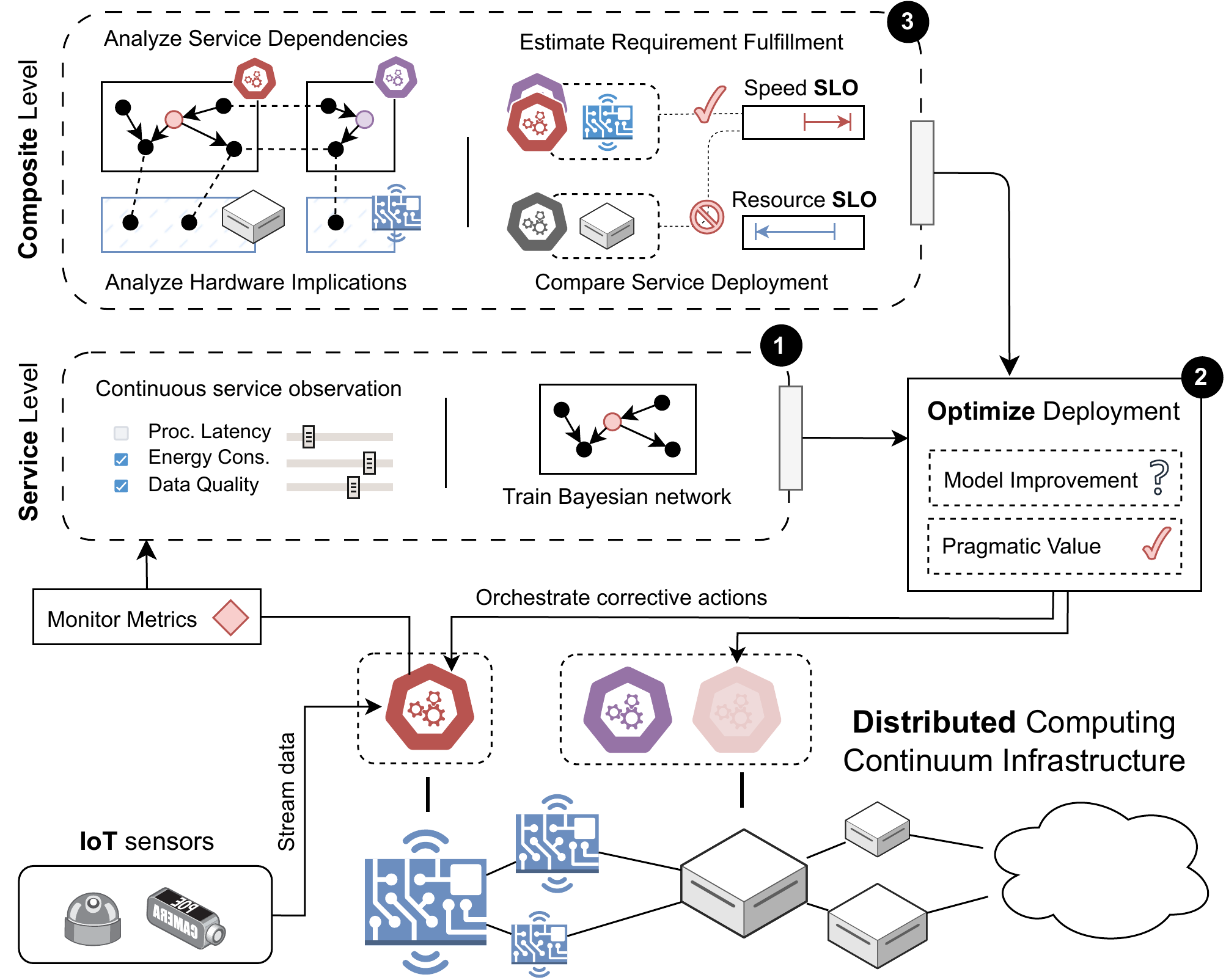}
    \caption{Three-level implementation for autonomous service orchestration in the Computing Continuum: \blackcircled{1} we interpret current service states (e.g., \textit{why} was an SLO violated) by training behavioral Markov blankets (MBs) from processing metrics; \blackcircled{2} agents continuously optimize the service operation according to their internal understanding of the environment (i.e., their MB) and current SLOs. Lastly, \blackcircled{3} to optimize SLO fulfillment throughout the CC, we compose MBs to quantify dependencies between services and hosting devices.}
    \label{fig:phd-contributions}
\end{figure*}

This section outlines a three-level implementation for autonomous service orchestration in CC systems (see Figure~\ref{fig:phd-contributions}) that follows the abstract solution design and its agentic lifecycles. First, we model dependencies between services, the processing environment, and SLO fulfillment, e.g., quantify the impact of GPU resources on \textit{energy efficiency} and \textit{latency}. Next, we update these models through a continuous action-perception cycle to keep them up-to-date with changing context; lastly, we compose models into larger structures that quantify dependencies between services, and their resource demand according to services' configuration.

\subsection{BEHAVIORAL MARKOV BLANKETS}

A MB marks the statistical boundary between an entity and its environment---this defines what it can directly observe and control, and separates it from whatever lies beyond these limits. We use this concept to model \textit{how} a service (or rather its SLOs) is affected by its direct observations and the actions it could take. For example, how much \textit{load} an AR application can take before dropping its \textit{performance}, and how lowering the \textit{rendering quality} can recover \textit{performance}. The precise relations and conditional dependencies between these variables are modeled through a Bayesian Network (BN); for instance, the BN could express a function of how the rendering \textit{quality} impacts service \textit{performance}, and given that we change the \textit{quality} to a certain level, what \textit{performance} to expect.

The MB contains all actions that an application or service can initiate itself; when choosing between actions, the BN can answer which action promises the highest SLO improvement. For instance, given that application \textit{performance} is declining, the choice might be between scaling up \textit{resources} or decreasing \textit{quality}. Conditioned on the current state---which tells that additional \textit{resources} are either expensive or depleted---it is possible to make an informed decision~\cite{sedlak_designing_2023_short}. This greatly improves applications' flexibility because individual services can act according to changing context.

However, relations and dependencies that make up the MB might either be unknown at design time or change dynamically; to that extent, we continuously update the MBs at runtime using observational data. 
In our AR overlay example, we monitor the application and collect operational metrics (e.g., rendering \textit{latency}) to evaluate SLOs fulfillment. 
Whenever SLOs change (e.g., AR application forced to save \textit{energy}), this merely changes the desired metrics distribution. 
These steps correspond to \blackcircled{1} in Figure~\ref{fig:phd-contributions}. 
%
%Validation of a smartphone-based tremor measurement tool for Parkinson’s disease
Thus, we analyze the precise impact of environmental dynamics and heterogeneous factors during runtime.

\subsection{ACTIVE INFERENCE CYCLE}

% By training their own behavioral MB, services become aware of their interplay with their surrounding. In particular, how to perceive their environment and enact on it without external instructions. For the opposite, imagine an application deployed on an Edge device in a traffic junction; if the service faces high demand at rush hours or poor quality during night, this needs to be resolved timely. However, delegating the decision-making to the Cloud requires sharing vast amounts of device states, which congests the network and slows down reaction, but also poses legal problems for privacy protection. Furthermore, IoT environments are highly dynamic, causing variable drifts, and consequently, inaccurate decision models. This motivates a decentralized solution for lifelong learning, running at the Edge.

% Using this BN, we can build an agent that monitors the service and recommends actions (e.g., scaling) that optimize the service quality. 
% For instance, the agent can approximate how much load the service can endure, and how a corrective action, e.g., decreasing \textit{video resolution}, impacts the service quality.

Equipped with their own behavioral MBs, smart services can perceive and act on their environment without external instructions. For instance, a service agent can approximate how much load a service can endure before dropping its quality, and take a corrective action, e.g., decreasing sensor \textit{resolution}, to restore service \textit{quality}. When processing sensor data on an Edge device---like movement data of habitants in a smart city---the demand spikes at rush hour and video quality drops at night; to ensure service quality, the agent must act timely. However, running decision-making in the Cloud requires sharing vast amounts of service and device states, which congests the network, slows down reaction, and poses legal problems for private data. Additionally, dynamic IoT environments are prone to variable drifts---leading to inaccurate models.
% This motivates a decentralized, lifelong learning approach at the Edge~\cite{danilenka_adaptive_2024}.

To ensure that corrective actions show the expected and desired outcome, we wrap a service and its decision model---the MB---into a continuous action-perception cycle, guided by AIF; notice how this implements the abstract control loop presented earlier. In each iteration, the service agent evaluates how accurately it predicted the effects of its last action and updates its MB. In more detail, this means adjusting the conditional variable dependencies according to new observations, for instance, because it overestimated the effect of adjusting the sensor \textit{resolution}. Afterwards, the agent chooses between actions that promise high information gain for improving the MB accuracy, or such that fulfill its preferences, e.g., low rendering latency.
This is shown by \blackcircled{2} in Figure~\ref{fig:phd-contributions}.

To allow agents persist over time, AIF aims to minimize \textit{free energy}~\cite{parr_active_2022}---an upper bound for uncertainty in a system. Originating from neuroscience, AIF can be used to guide the behavior of humans and service agents alike. Within the control loop, our service agent aims to develop an accurate (i.e., unsurprising) model of the environment, and use it to ensure each component's SLOs. This sets AIF apart from pure RL, because AIF prioritizes model accuracy over solely maximizing cumulative reward~\cite{tschantz_reinforcement_2020}. By integrating AIF into service orchestration, 
we ensure that software components have an accurate understanding of how their local actions can optimize SLO fulfillment.
% we continuously satisfy services' internal requirements and ensure that corrective actions rely on accurate assumptions.

\subsection{COLLABORATIVE ORCHESTRATION}

Every computing service has specific demands: video processing services might favor infrastructure with integrated GPUs, or force other services to change their configuration, e.g., downstream services might depend on an upstream service to improve its quality. Other services might be less computationally intense, but require low-latency placement close to an IoT data source. If we deploy services without considering these constraints, it inevitably leads to significant SLO violations. Often an optimal solution is non-trivial: for example, if we have three services that require GPU support and only resources to accommodate one, how to allocate the resources?
This shows that service orchestration in resource-restricted environments cannot be dictated by a single actor, but requires their collaboration to find a global and fair optimum.

To support decentralized and collaborative actuation, we analyze the dependencies between services and hosting devices, more precisely, between the variables located in their MBs. Thus, we explore whether actions have an impact on nearby entities, e.g., if it affects the performance and resource consumption of downstream services, or any service co-located on the same device. 
% This allows us to build one composite model and infer expected quality of downstream services according to the configurations of upstream services and the allocated hardware resources.
Consequently, individual components become aware how local actions (e.g., reconfiguring a service) impact the entire system; this is shown by \blackcircled{3} in Figure~\ref{fig:phd-contributions}. 
Whenever SLOs change throughout the application, e.g., the desired quality of one service within a pipeline, the upstream services will explore their action space to improve downstream SLO fulfillment.
Again, this is a powerful upgrade for the autonomy of computing services because they can jointly find and execute an optimal corrective action.

% \tmi{
\subsection{PRELIMINARY RESULTS}

To ensure the soundness of our three-layer methodology, we evaluated the individual parts through a series of experiments~\cite{sedlak_designing_2023_short,sedlak_equilibrium_2024,sedlak_markov_2024_short}. We summarize the key findings below. In particular, we showed that: (1) training BN and extracting the MB for autonomous processing services
% ---as part of the AIF action-perception cycle---
did not introduce a critical overhead for common edge devices (e.g., Nvidia Jetson), (2) MBs allowed services to ensure SLO fulfillment by taking optimal actions; at the same time, the BN structure allowed empirically interpreting and verifying these actions, and (3) for composite microservice applications, the MBs of individual services could be equally composed to analyze dependencies between services and optimize global SLO fulfillment. 
% Naturally, early studies had to carry out experiments with reduced scope, so that the effectiveness of our approach still has to be proved in larger environments. Nevertheless, we believe that the proposed conceptual model has strong potential to deliver significant benefits in complex and heterogeneous CC environments. 

\subsection{SUITABILITY OF THIS IMPLEMENTATION}

Considering the extension of structural problems in the CC, no single solution can wrap all---this also shows shortcomings of existing works~\cite{lan_sla-orecs_2024,kochovski_smart_2020,piaseczny_rccda_2025}.
Rather, solutions must be modular and evolve organically. In this section, we discuss aspects covered by our preliminary solution, and critical open research gaps. 

Starting from the \textbf{domain hypotheses}, our approach considers heterogeneity (hardware capacity, energy consumption, network constraints) in the processing environment (D.1) and models dynamic environmental conditions as factors inside the MB (D.2); however, it focuses on fast-paced fluctuations, leaving a gap for hierarchical solutions operating at different time scales. The largest gap stems from the difficulty and operational overhead of evaluating orchestration mechanisms in large-scale environments (D.4).

Our approach uses local observations to quantify their implication to global objectives (D.3); however, connecting to the \textbf{application hypotheses}, there is a large gap for multi-vendor solutions (A.2) to exchange observations and agree on joint policies. Still, our approach makes use of clear monitoring and action interfaces between services (A.4), which it uses to quantify dependencies between services, their SLOs (A.3), and the processing hardware.
Thus, the presented AIF agents follow the \textbf{invariant} action-perception cycle and the stochastic dynamics of the processing environment (M.3).
While we evaluated different \textbf{variants} for agents' algorithms and configurations, it remains to analyze different hierarchical and evolving architectures.
% and different internal belief models for agents.
% \ins{map/bridge to the challenges. Also, might mention that we test various \textbf{variants} (e.g., algorithms etc), while strictly following the \textbf{invariant} agent cycle.}

\section{RESEARCH CHALLENGES}

% The presented solution aims to cover large parts of the inherent problems in CC systems. Nevertheless, it leaves numerous open research challenges---we argue that many of them are structural and generally impede research on service orchestration. In the following, we recapitulate the most critical ones and formulate three research challenges that promise structural improvement for services computing.
% %
% Most importantly, we lack: (1) large-scale, standardized simulation environments that quickly allow testing hypothesis and comparing solutions, (2) continuous adaptation of explainable AI solutions, and (3) seamless infrastructure composition that allows clients to use infrastructure from multiple vendors or individuals.

The presented implementation covers large parts of the inherent problems in CC systems, yet it leaves numerous open research challenges---we argue that many of them are structural and generally obstruct research on service orchestration. Below, we recapitulate the most critical gaps and formulate three challenges for accelerating research on service computing.
Most importantly, we lack: (1) large-scale, standardized simulation/emulation environments for quickly testing hypotheses and comparing solutions, (2) context-aware and resilient mechanisms for updating and invoking orchestration mechanisms, and (3) seamless infrastructure composition that allows clients to combine devices from multiple vendors and their individual repertoire.
% %
% To that extent, we formulate three research challenges that promise structural improvement---not only for preliminary solution---but services computing in general.

% To scale our service orchestration methodologies to larger setups, we face numerous technical challenges; we argue that many of them are structural in the field of service orchestration and generally slow down research progress. 
% %
% \tmi{Most notably, we found that service orchestration could strongly benefit from simulation and evaluation environments that allow simple training, testing, and comparison of different orchestration algorithms in standardized problem instances. Also, the CC is envisioned as an overarching platform that comprises different vendors; as such, there is a large gap for orchestration mechanisms that support multi-vendor setups, or allow clients to add personal devices, over which the exhibit maximum sovereignty.}
%
% In the following, we will explain these challenges in more detail and underline how they are both relevant to accelerate general progress in service-oriented computing, and at the same time, help us scale and improve the presented orchestration methodologies.
%
% While the list of presented challenges is not exhaustive, we selected them according to their general impact on service-oriented computing and on the presented approach for autonomous service orchestration.

% \ins{VCP: Need to make sure that all challenges are at the same level of a taxonomy.}

\subsection{SYSTEM \& ACTION SIMULATION}

\begin{figure}
% \vspace{-8pt}
    \centering
    \includegraphics[width=1.02\linewidth]{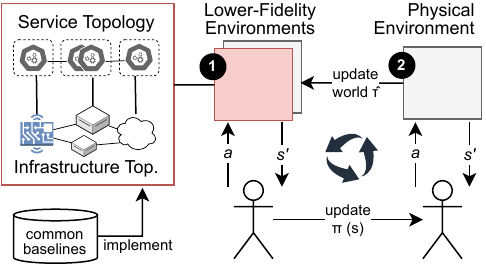}
    % \vspace{1pt}
    \caption{Missing an ecosystem for \blackcircled{1} pretraining agents in lower-fidelity environments (e.g., simulation) implementing common baselines and infrastructure; \blackcircled{2} agents are deployed in physical devices and pretraining environments improve through real-world feedback.}
    \label{fig:challenge-simulation}
\end{figure}

Empirical evaluations are a standard way to demonstrate technical soundness of hypotheses. However, scaling them to larger environments requires substantial on-demand resources, which are owned by few large-scale companies. For researchers outside this league, the cost of physical experiments (e.g., setup time or hardware expenses) can quickly outweigh benefits~\cite{bowden_designing_2023}. At the same time, extracting and implementing baselines from reference literature is often a tedious and time-consuming task. To lower the resource barrier and enable continuous integration with empirical or emulated environments, we emphasize the following two-stage challenge alongside Figure~\ref{fig:challenge-simulation}.

\subsubsection{Baseline Standardization}

The rise of RL has been supported by an entire ecosystem of tools that allow developers to quickly test their ideas on standardized problems (e.g., \textit{OpenAI Gymnasium}\footnote{https://gymnasium.farama.org/index.html}) or compare them with contemporary algorithms (e.g., \textit{Stable Baselines}\footnote{https://github.com/DLR-RM/stable-baselines3}--SB3).
While service computing also knows standardized problem instances, such as the \textit{DeathStarBench} suite,\footnote{https://github.com/delimitrou/DeathStarBench} running its manifold of microservices requires sophisticated hardware and slows down ML training, since results can only be obtained in real time.
Moreover, although baselines such as SB3 are easily accessible, they typically contain generic versions of algorithms that can easily be outperformed by more optimized implementations~\cite{engstrom_implementation_2020}, thus limiting the rigor and comparability of experimental results.

To simplify the training, testing, and deployment of ML solutions in service computing, we identify a substantial gap in lower-fidelity environments (e.g., simulation) that allow researchers and developers to (1) quickly obtain results for training and testing, and (2) compare their solutions with state-of-the-art algorithms. Addressing this challenge would simplify preparation and execution of experiments---lowering the entry barrier for R\&D and SMEs---and improve results rigor through standardized baseline comparison.

\subsubsection{Digital Twinning}

Simple simulation environments, such as \textit{OpenAI Gym}, provide hands-on ways to engage with ML technologies, but offer limited practical value for pretraining agents on real-world problems. More sophisticated simulators, such as NVIDIA Isaac\footnote{https://developer.nvidia.com/isaac}, combine physics engines and real-world benchmarks to create more realistic environments. This mechanism—integrating benchmarks and real-world feedback—is a core feature of digital twins and may form the backbone of self-evolving simulators~\cite{pretel_active_2025}.

A complete ecosystem for accurate benchmarking of CC applications could combine different \textit{fidelity levels}, where simulated results are gradually validated in emulated platforms (e.g., iContinuum~\cite{akbari_icontinuum_2024}) or physical testbeds. This ecosystem could also help replicate realistic system operations from physical setups down to simulation environments; if simulations face realistic conditions (e.g., node failures), it also improves their rigor. Not only would it be costly to train the agent entirely on real node failures, but these negative situations also occur rarely, so the agents can use the lower-fidelity levels to explore these situations thoroughly, and use higher-fidelity environments for fine-tuning.

\subsection{MODEL UPDATE \& INFERENCE}

\begin{figure}
% \vspace{-7pt}
    \centering
    \includegraphics[width=1.02\linewidth]{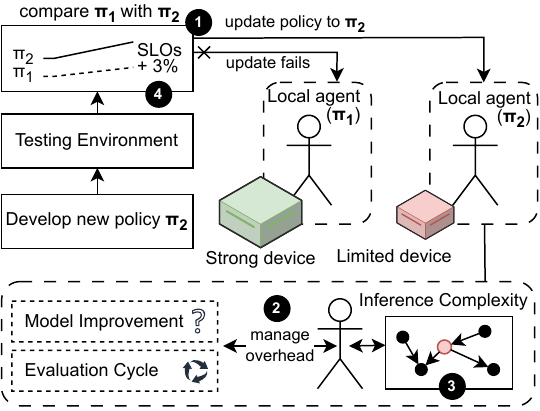}
    \caption{Updates to orchestration policies (e.g., from $\pi_1$ to $\pi_2$) must \blackcircled{1} balance the impact to performance, and tolerate inconsistent model rollout; \blackcircled{2} agents must minimize their execution overhead according to the context, and keep the user in the loop to  \blackcircled{3} develop new metrics and \blackcircled{4} align with user intents.}
    \label{fig:challenge-models}
% \vspace{-4pt}
\end{figure}

Service orchestration requires accurate policies, e.g., for fine-grained scaling of resources or replicas. However, as the environment changes, policies quickly become outdated or fail to capture the new context, e.g., as clients change SLOs. Thus, scaling agents might fail to predict the next state and reward accurately, or exceed the available timeframe for inferring a policy. This is particularly critical for increasingly smaller devices and autonomous components that might not possess sufficient resources to run sophisticated mechanisms for finding and updating policies. We summarize four problems alongside Figure~\ref{fig:challenge-models}: first, continuously update policies and rollout changes despite dynamic disruptions, and second, manage agents' execution overhead on devices with limited resources. Third, we stress new new performance metrics to identify bottlenecks and disruptions within the environment; and fourth, require a human-in-the-loop to ensure correctness.

\subsubsection{Continuous Learning at Scale}

To maintain their perception, predictions, and actuation accurate, autonomous agents must adapt to new data---a process known as continuous learning. Given the extension and highly-distributed nature of CC applications, this requires communicating and coordinating policy updates at unprecedented scale.
Consider alone the impact of inconsistent or brittle network conditions during model rollout: the outcome will be agents with poorly synchronized models, causing undesired effects, like unfair resource allocations among tenants. From a networking perspective, we argue that dynamic disruptions cannot be completely prevented, and hence, agents require robust orchestration mechanisms to (co-)operate gracefully despite diverging policy versions. Simulating failures for improving the training stability can be supported by common tools, like Chaos Mesh\footnote{https://github.com/chaos-mesh/chaos-mesh}.

At the same time, the overhead of continuously integrating model updates must be weighed against the accuracy gained in return, and the cold-start introduced by any model update. This creates an optimization problem in which agents' hyperparameters (e.g., update rate or accuracy thresholds) need to be adjusted at numerous points in the architecture. However, current ML solutions often rely on static or central hyperparameter optimization---as common in Federated Learning (FL)~\cite{danilenka_adaptive_2024}---without the options to dynamically adapt and reconfigure the learning process.
This solicits mechanisms that align the update rate with hardware limitations~\cite{piaseczny_rccda_2025}, however, the challenge is integrating such context-aware approaches at larger scale, and performing dynamic hyperparameter optimizations at increasingly smaller components.

\subsubsection{Context-aware Orchestration Quality}

Time-critical domains, like robotic control~\cite{engstrom_implementation_2020}, frequently require complex operations on sensory data. While the CC---and in particular Edge computing---can deploy services in close vicinity, this exposes services to dynamic environmental conditions, like fluctuating load and limited resources.
Under these circumstances, scaling agents must equally be able to orchestrate services under strict latency requirements. For example, an AR-supported robot navigating a smart city might offload and scale nearby services in real-time.
%
% Despite these conditions, scaling agents must ensure latency constraints for inferred actions; at the same time, they must not impede other services running in the same device, but provide the highest quality with given resources.
%
To not impede other services running on these devices, scaling agents must consider their own computation overhead, the effects on co-located applications, and deduce hardware constraints for inference; for example, prioritize fast autoscaling cycles over their high accuracy. Depending on this context, scaling agents could adjust their internal policy calculation, e.g., by switching between ML models~\cite{pretel_active_2025}, or early-exiting NNs~\cite{chen_ceed_2025}. 
% The challenge, however, is finding such adaptation mechanisms for arbitrary ML services and integrate them in continuous feedback cycles.
%
% Considering the preliminary solution described in Fig.~\ref{fig:phd-contributions}, this might require to compose smaller/larger neighborhoods of services to determine the optimal action under less/more knowledge.

\subsubsection{Novel Performance Metrics}

Understanding a system often depends on interpreting metrics that provide insight into hidden states~\cite{sedlak_equilibrium_2024}. As CC environments become increasingly dynamic and complex—featuring heterogeneous devices, fluctuating resource availability, and frequent mobility—there is a growing need for context-aware metrics that reflect operating conditions.
% ~\cite{donta_performance_2025}.
For example, \textit{next-level metrics} could account for environmental volatility, multi-layer coordination costs, or a system’s ability to stabilize performance under uncertainty. Creating such metrics may require capturing new states or combining existing ones.
Incorporating these metrics into simulation and testing frameworks would enable more realistic assessments of orchestration, scheduling, and adaptation mechanisms.

\subsubsection{Intent Control \& Human-in-the-Loop}

Agents can be trained to ensure SLO fulfillment, e.g., through RL and control-loop systems~\cite{lan_sla-orecs_2024}. This, however, does not guarantee that system behavior actually aligns with strategic goals of human operators. To ensure a match, orchestration must shift from rigid SLO definitions to intent-based control: operators define high-level intents, which are continuously diffused to lower-level SLOs and configurations. In this loop, Large Language Models (LLMs) could serve as critical intermediaries, translating high-level human intents (e.g., "prioritize low latency for emergency streams") into scaling actions (e.g., offload services to make space for an emergency service)~\cite{su_llm-driven_2026}. This would democratize access to complex infrastructure, as non technical clients (e.g., city planners) can deploy applications and rely on automatic orchestration according to their intents.

% Human-in-the-loop \& intent-based control shifts service orchestration from low-level, fully autonomous decisions to human-guided, goal-driven control, where operators express what they want, not how to do it, and the system adapts accordingly. This is an important challenge which is missing in my opinion. It also complements Digital Twins and Novel Metrics extremely well.

\subsection{INTERPLAY OF INFRASTRUCTURE}

\begin{figure}
% \vspace{-8pt}
    \centering
    \includegraphics[width=1.02\linewidth]{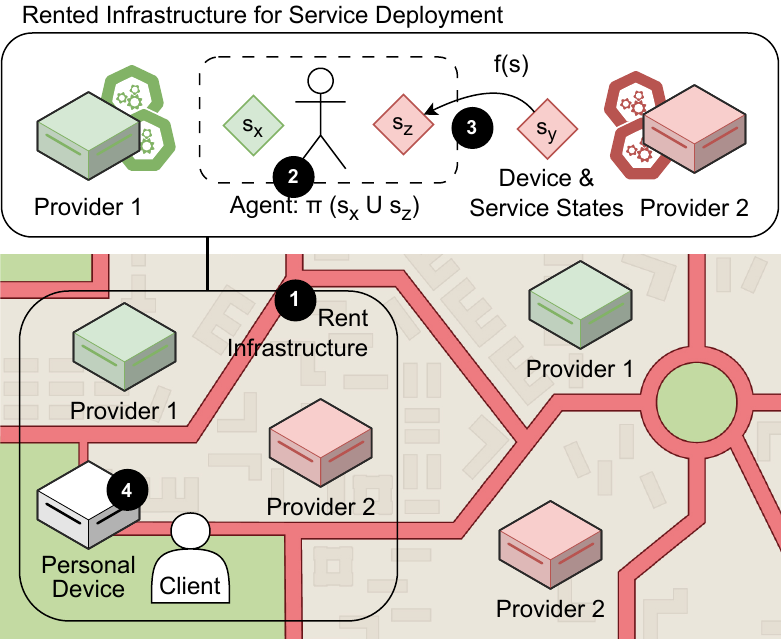}
    \caption{Client wants to \blackcircled{1} rent and combine the nearest infrastructure from multiple vendors; agents must \blackcircled{2} orchestrate and cooperate towards global objective, \blackcircled{3} despite limited visibility; ideally, this allows \blackcircled{4} seamless interplay with client-owned devices.}
    \label{fig:challenge-infrastructure}
\end{figure}

A vision of CC systems is to combine infrastructure layers into a single platform, where services are optimized autonomously, encapsulated from the application provider. However, looking at today's provider landscape, e.g., for Cloud infrastructure, there is no interplay between providers~\cite{saxena_survey_2021}: clients must weigh between different providers and their subset of the overall infrastructure.
Another promise of the CC—or Edge computing—is processing confidential data close to the data source. While Edge devices are integrated into providers' portfolios, like AWS Outposts,\footnote{https://aws.amazon.com/edge/services/}
 data is still processed outside the customer's sovereignty. To support the collaboration of infrastructure providers and empower citizens to take control of personal data, we formulate four challenges alongside Figure~\ref{fig:challenge-infrastructure}.

\subsubsection{Multi-Vendor Governance}

Collaboration between different infrastructure providers is a necessity for optimizing the customer experience; ideally, service consumers could process data on the nearest device---despite provider boundaries. The fragmented infrastructure landscape is primarily caused by business interests, but regardless, brings numerous technical challenges, such as distributing the governance between multiple vendors. Consider the following example: we scale an application horizontally and place replicas on devices of operator $A$, and also operator $B$; how would these operators divide the responsibility to ensure the application goals, or more critically, recompense the client in case of SLA violations? 

To tackle this challenge, we see a design-time and runtime part to it: during \textbf{design time}, agreements must be formed between infrastructure providers, and between providers and clients. This process can be supported through smart contracts~\cite{kochovski_smart_2020} and might lead to a multitude of SLAs. However, as SLOs are violated, there is a gap for accounting mechanisms that identify the partial faults of providers and automatically recompense clients accordingly. During \textbf{runtime}, the service deployment must be optimized across vendors---an upgrade to existing single-vendor mechanisms~\cite{lan_sla-orecs_2024}. 
% Complementary to the presented methodology, re-orchestrate services of services could also be triggered by dynamic software pricing or clients' depleting budget.

% ~

% Motivated by the need for collaboration among different infrastructure providers, the development of CC management systems will likely take the form of \emph{decentralized applications}, where individual agents face partial observability and partial cooperation.

\subsubsection{Partial observability}

CC systems use infrastructure from multiple vendors that operate across different data jurisdictions. The constraints imposed by these jurisdictions, together with vendors' privacy regulations, restrict data monitoring and remote intervention capabilities. These limitations motivate research into multi-agent algorithms capable of performing effectively even when information sharing is only possible between some jurisdictions, vendors, or agents, and when data sharing must comply with specific privacy constraints.  

\subsubsection{Partial cooperation}

Even when vendors act honestly and comply with their individual SLAs, a multi-vendor setup can still drift into non-cooperative behavior. This resembles multiple musicians playing their parts perfectly yet producing a dissonant ensemble. In practice, vendors may prioritize actions that safeguard their own SLA compliance rather than collaborating with others. Under tight conditions, this can lead to \emph{handoff timing optimizations} (accelerating local processing in ways that shift load to downstream vendors), or \emph{selective throttling} (limiting request or resource usage for certain clients).
Hence, we must detect non-cooperative vendor behavior without breaking their autonomy.

Altogether, the CC demands decentralized orchestration mechanisms that support: partial visibility, partial cooperation, and continuous updates. Combining these features, however, easily results in oscillating actions and cascading failures. Therefore, risk management must equally be decentralized so that individual services are always accountable for failures they cause. This incentivizes service providers to develop a clear model of the quality and reliability they can provide; otherwise, SLA violations will prove costly and service consumers will avoid using the service.

\subsubsection{Democratizing Infrastructure}

The CC poses the unique opportunity to allow service users to integrate their own infrastructure into a larger computing platform. First, this would allow clients to constrain confidential operations to local or neighboring device, which they fully trust and over which they exhibit a certain sovereignty. Second, by integrating their devices into the CC, clients would benefit from higher availability and fault tolerance---like offloading computations to remote nodes if they exceed local capabilities.

To empower service consumers in controlling the flow and processing of data, there is active research on:
% decentralizing the web and its services, particularly on the 
access and governance of personal data~\cite{verborgh_re-decentralizing_2023}, and creation of trusted execution environments~\cite{munoz_survey_2023}. A logical next step would be to integrate these mechanisms into the CC, and allow individuals to add personal infrastructure.
% , which might increase heterogeneity even more. 
This faces numerous challenges, most critically, ensuring the security of consumer devices, but also service orchestration. Namely, we lack mechanisms to scale services between trusted environments, and business models that reward device owners for providing their own personal infrastructure.
% Notably, these challenges can benefit largely from solutions for multi-vendor setups.
% Considering the existing support for heterogeneous devices, research will mostly have to focus on establishing trust and 

\section{CONCLUSION}

This article provides a structural perspective into the Computing Continuum (CC) and the challenges that its heterogeneous and dynamic infrastructure raises for service orchestration. To guide research in that domain, we first envision an autonomous solution for service adaptation, and second, show how Active Inference (AIF)---a concept from neuroscience---could fill this gap. Following an agentic lifecycle, services can interpret their context, maintain accurate internal models, and reason about interdependencies with other services. While our implementation addresses fundamental problems---like heterogeneous and dynamic operating conditions---numerous aspects remain unaddressed by contemporary research. To bridge this gap, we formulate three research areas that aim to simplify the execution and interpretation of experiments, the continuous and efficient updates of AI models, and the interplay of infrastructure from multiple providers and clients.
We invite other researchers to break down these structural barriers together and accelerate research on service orchestration in the CC.

\section{ACKNOWLEDGMENTS}
\vspace{2pt}

This work is supported by CNS2023-144359 financed by MICIU/AEI/10.13039/501100011033 and the European Union NextGenerationEU/PRTR.

\bibliographystyle{IEEEtran}
\bibliography{short,Boris}
% \bibliography{short}

% \newpage
\input{x_biography}

\end{document}

%% file: x_biography.tex
\begin{IEEEbiography}{Boris Sedlak}{\,}  is Postdoctoral researcher at the Engineering department of UPF, Barcelona, in the Distributed Intelligence and Systems-Engineering Lab (DISL).
% prior to that, he received his Ph.D. from TU Wien.
Contact him at \texttt{boris.sedlak@upf.edu}. \vspace*{2pt}
\end{IEEEbiography}

\begin{IEEEbiography}{Víctor Casamayor Pujol}{\,} is a Tenure track professor at the Engineering department of UPF, Barcelona, in the Distributed Intelligence and Systems-Engineering Lab (DISL). 
Contact him at \texttt{victor.casamayor@upf.edu}. \vspace*{2pt}
\end{IEEEbiography}

\begin{IEEEbiography}{Ildefons Magrans de Abril}{\,}  is currently working as a Postdoctoral researcher at the Engineering department of UPF, Barcelona, in the Distributed Intelligence and Systems-Engineering Lab (DISL). 
Contact him at \texttt{ildefons.magrans@upf.edu}. \vspace*{2pt}
\end{IEEEbiography}

\begin{IEEEbiography}{Praveen Kumar Donta}{\,} (SM'22) is Associate Professor (Docent) at the Department of Computer and Systems Sciences, Stockholm University, Sweden. %He received his Ph.D. from the Indian Institute of Technology (Indian School of Mines), Dhanbad, in the Department of Computer Science \& Engineering in June 2021. He is Senior Member of IEEE Computer and Communications Societies, and Professional Member of ACM. 
His research includes learning in Distributed Continuum Systems. Contact him at \texttt{praveen@dsv.su.se}. \vspace*{2pt}
\end{IEEEbiography}

\begin{IEEEbiography}{Adel N. Toosi}{\,}
is an Associate Professor of Computer Systems and Director of the \textit{Dis}tributed Systems and \textit{Net}work Applications Lab (\textit{DisNet} Lab) at the University of Melbourne, Australia. Contact him at \texttt{adel.toosi@unimelb.edu.au}.
\vspace*{2pt}
\end{IEEEbiography}

\begin{IEEEbiography}{Schahram Dustdar}{\,}  is Full Professor of Computer science heading the Research Division of Distributed Systems at the TU Wien, Austria. He is also part-time research professor at ICREA | UPF Barcelona. Contact him at \texttt{dustdar@dsg.tuwien.ac.at}. 
% \vspace*{2pt}
\end{IEEEbiography}